\documentclass[prb,aps,preprint,showpacs,superscriptaddress]{revtex4}

\usepackage{graphicx}

\begin{document}

\title{Single-ion anisotropy, Dzyaloshinskii-Moriya interaction and negative magnetoresistance 
of the spin-1/2 pyrochlores R$_2$V$_2$O$_7$}

  \author{H. J. Xiang} 
  \affiliation{Key Laboratory of Computational Physical Sciences (Ministry of Education), and Department of Physics, Fudan
    University, Shanghai 200433, P. R. China}

  \author{E. J. Kan}
  \affiliation{Department of Chemistry, North Carolina State
    University, Raleigh, North Carolina 27695-8204, USA}

  \author{M.-H. Whangbo}
  \affiliation{Department of Chemistry, North Carolina State
    University, Raleigh, North Carolina 27695-8204, USA}

  \author{C. Lee}
  \affiliation{Department of Chemistry, North Carolina State
    University, Raleigh, North Carolina 27695-8204, USA}

  \author{Su-Huai Wei}
  \affiliation{National Renewable Energy Laboratory, Golden, Colorado 80401, USA}

  \author{X. G. Gong}
  \affiliation{Key Laboratory of Computational Physical Sciences
    (Ministry of Education), and Department of Physics, Fudan
    University, Shanghai 200433, P. R. China}

\date{\today}

\begin{abstract}
The electronic and magnetic properties of spin-1/2 pyrochlores R$_2$V$_2$O$_7$ were
investigated on the basis of density-functional calculations. 
Contrary to the common belief, the spin-1/2 $V^{4+}$ ions are found to have a substantial 
easy-axis single-ion anisotropy. The $|D/J|$ ratio deduced from the magnon quantum Hall effect 
of Lu$_2$V$_2$O$_7$, where $J$ is the nearest-neighbor spin exchange
and $D$ is the Dzyaloshinskii-Moriya parameter, 
is much greater than the value estimated from our calculations (i.e., 0.32 vs. 0.05). 
We show that this discrepancy is due to the neglect of the single-ion anisotropy of the 
V$^{4+}$ ions, and the negative magnetoresistance observed for R$_2$V$_2$O$_7$ 
arises from a new mechanism.

\end{abstract}

% 75.10.-b
%Magnetic anisotropy, 75.30.Gw
%71.15.Rf Relativistic effects
%71.20.-b Electron density of states and band structure of crystalline solids

\maketitle
%introduction

During the past two decades the magnetic properties of pyrochlore
oxides of the type R$_2^{3+}$M$_2^{4+}$O$_7$
(R = rare-earth; M = transition metal)
have been extensively studied \cite{Gardner2010,Blaha2004},
due largely to the spin frustration in the
pyrochlore lattice that results when the nearest-neighbor (NN) spin
exchange is antiferromagnetic
(AFM) \cite{Gardner2010}. The vanadate pyrochlores 
R$_2$V$_2$O$_7$ (R = Lu, Yb, Tm, Y) are unique because
they are ferromagnetic insulators
\cite{Shamoto2001,Shamoto2002,Knoke2007,Ichikawa2005,Kiyama2006,Miyahara2007}
contrary to the observation that ferromagnetism leads usually to metallic character.
Furthermore, Lu$_2$V$_2$O$_7$ exhibits a negative magnetoresistance (MR) as high as
50\% just above the Curie temperature $T_C = 73 K$ under magnetic field of 5 T
\cite{Zhou2008}. Recently, Lu$_2$V$_2$O$_7$ is found to exhibit a magnon Hall effect
(i.e., the anomalous thermal Hall effect caused by spin excitations) \cite{Onose2010}.
The explanations presented for these observations raise 
fundamental questions. Namely, 
the MR of Lu$_2$V$_2$O$_7$ was suggested to be caused 
by polaron mediation as found in Tl$_2$Mn$_2$O$_7$ \cite{Zhou2008}. However, this
possibility seems remote, because the diffuse $6s$ orbital of the
Tl$^{3+}$ ion is believed to assist the polaron formation in Tl$_2$Mn$_2$O$_7$ whereas the
Lu$^{3+}$ ion in Lu$_2$V$_2$O$_7$ does not have such an extended $s$ orbital. In the
magnon Hall effect \cite{Katsura2010}, the Dzyaloshinskii-Moriya (DM) interaction is considered to play the role
of the vector potential as in the intrinsic anomalous Hall effect in metallic ferromagnets. In
the fitting analysis of their experimental data, 
Onose {\it et al.} found it necessary to use the $|D/J|$ ratio of 0.32.
This ratio is unusually large because the DM interaction is
a consequence of spin-orbit coupling (SOC) so that the typical $|D/J|$
ratio is expected to be smaller than 0.1. Therefore, it is important to quantify 
the magnitude of the DM term.
In this Letter, we probe these questions by studying the electronic
and magnetic properties of R$_2$V$_2$O$_7$
on the basis of density functional calculations to find that 
the spin-1/2 $V^{4+}$ ions have a substantial 
easy-axis single-ion anisotropy contrary to the common belief, the neglect of 
this anisotropy can lead to an unusually large  $|D/J|$ ratio, and 
R$_2$V$_2$O$_7$ exhibits a new type of negative MR mechanism.

%\clearpage

Our calculations are based on the density functional theory (DFT) plus the on-site
repulsion (U) method \cite{Liechtenstein1995} (DFT$+$U) within the generalized gradient
approximation  \cite{Perdew1996} on the basis of the projector augmented
wave method \cite{PAW} encoded in the Vienna ab initio simulation
package \cite{VASP}. The plane-wave cutoff energy was set to
400 eV. Careful convergence test were performed and the total energy
was converged to $10^{-6}$ eV.
In the following, we report results obtained with U $=$ 3 eV and J $=$ 1 eV on V,
but the dependence of our results on U and J will be also discussed.
It was found  \cite{Haghighirad2008} that Y$_2$V$_2$O$_7$ has  magnetic
properties very similar to those of Lu$_2$V$_2$O$_7$.
Thus, our calculations focused mainly on Y$_2$V$_2$O$_7$ using its
experimental structure \cite{Haghighirad2008}.
The structural optimization was
found to have a negligible effect on its magnetic properties.

\begin{figure}
  \includegraphics[width=7.0cm]{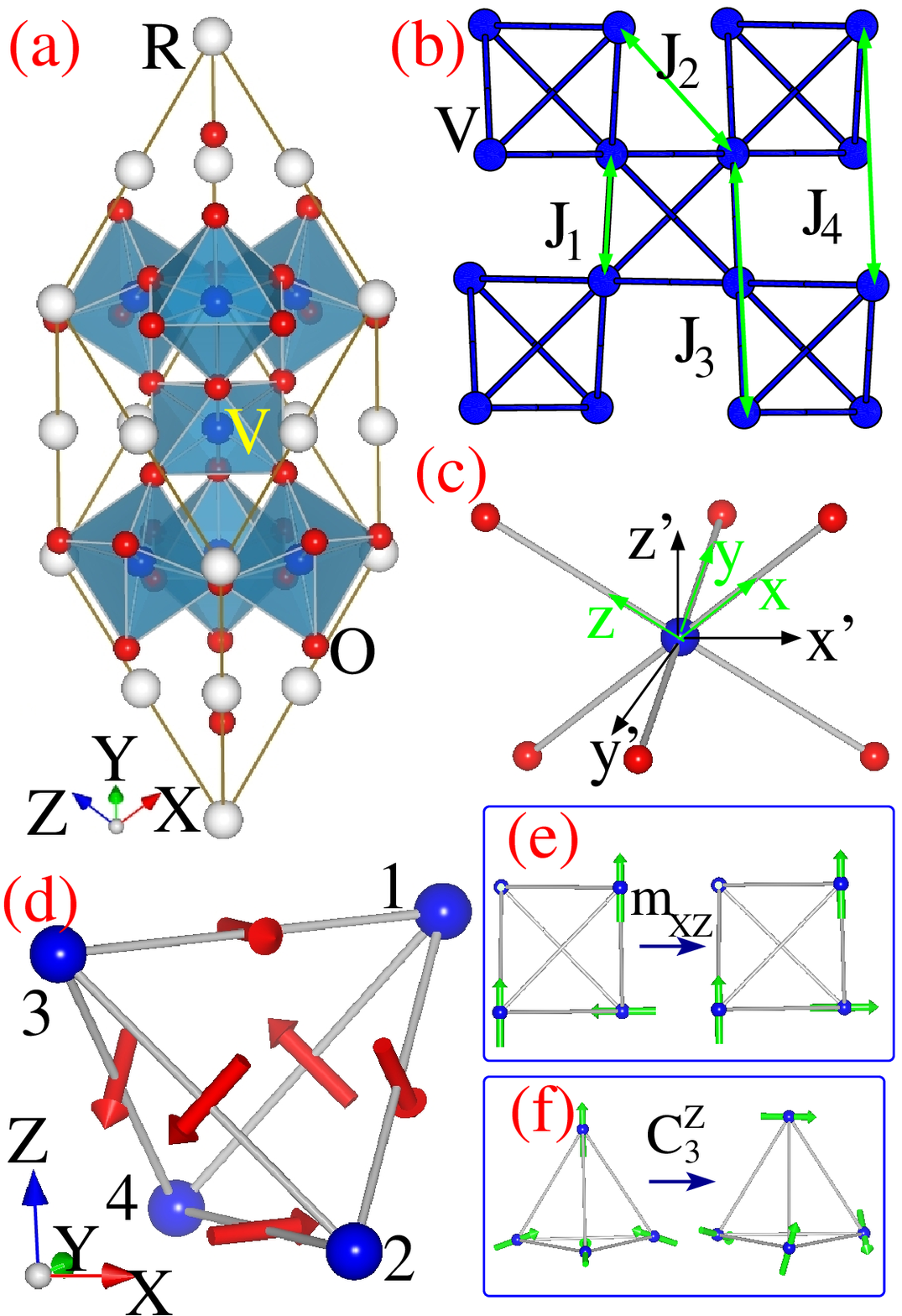}
  \caption{(a) The crystal structure of R$_2$V$_2$O$_7$. The global
    coordinate system $XYZ$ is indicated. (b) The spin
    exchange paths between V$^{4+}$ ions. (c)
    The two local coordinate systems used for
    the ideal VO$_6$ octahedron. The
    $xyz$ coordinate system is defined for the ideal VO$_6$ octahedron
    with the $x$, $y$ and $z$ axes
    taken along the V-O bonds. In the $x'y'z'$ coordinate system, the
    $z'$ axis is taken
    along one three-fold rotational axis of the ideal VO$_6$
    octahedron. In R$_2$V$_2$O$_7$ each
    VO$_6$ octahedron is axially compressed slightly with only one
    three-fold rotational axis.
    (d) The DM vectors
    ${\mathbf D_ij}$ with $i<j$ of the
    V tetrahedron, where $i$ and $j$ denote the V site labels. (e) The
    two spin configurations used to extract the DM parameter. (f) The
  two spin configurations used to extract the single-ion anisotropy
  parameter. }
  \label{fig1}
\end{figure}

In the V$_2$O$_7$ framework of Y$_2$V$_2$O$_7$ (space group $Fd\bar{3}m$), 
the VO$_6$ octahedron containing V$^{4+}$
 ($3d^1$, $S=1/2$) ions share their corners such that the V$^{4+}$ ions form corner-sharing
 V$_4$ tetrahedrons [see Fig.~\ref{fig1}(a)].
%% %The localized $S= 1/2$ spins at the V$^{4+}$ sites order ferromagnetically 
below $T_C \sim  70 $ K.
 The VO$_6$ octahedron of R$_2$V$_2$O$_7$ are slightly distorted (i.e., 
axially-compressed slightly)  from the ideal shape.
%so that the triply degenerate $t_{2g}$ level
%is split into a nondegenerate $a_{1g}$ and doubly degenerate $e'_g$
%levels.
Under the trigonal crystal field, the $t_{2g}$
states are split into the lowest $a_{1g}$ state $|0\rangle$ and
two $e'_g$ states $|+\rangle$ and $|-\rangle$ \cite{Miyahara2007}:
\begin{equation}
\begin{array}{lll}
  |0\rangle&=&1/\sqrt{3} (d_{xy}+d_{yz}+d_{xz})\\
  |+\rangle&=&-1/\sqrt{3} (d_{xy}+e^{2\pi i/3} d_{yz}+ e^{-2\pi i/3}d_{x} ) \\
  |-\rangle&=&1/\sqrt{3} (d_{xy}+e^{-2\pi i/3} d_{yz}+ e^{2\pi i/3}d_{xz} ),
\end{array}
\end{equation}
where the local coordinate system of a perfect VO$_6$ octahedron was
adapted [see Fig.~\ref{fig1}(c)].
Note that the higher $e_g$ states $d_{x^2-y^2}$ and $d_{z^2}$
remain degenerate. The recent
 polarized neutron diffraction study \cite{Ichikawa2005} showed that, at each V site
 of a given V$_4$ tetrahedron, only the $a_{1g}$ level is occupied by an electron.

%\clearpage

\begin{figure}
  \includegraphics[width=10.5cm]{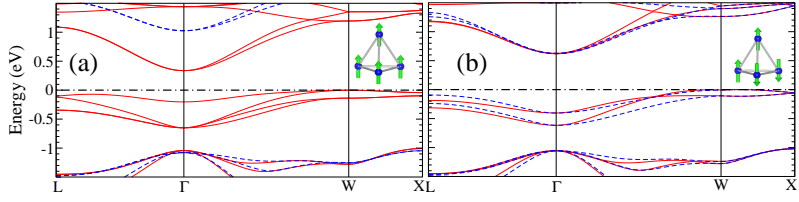}
  \caption{(color online) (a) The band structure of the FM state from the
    DFT$+$U calculation. (b) The band structure of the AFM state (with two up- and two down-spins
    in each V$_4$ tetrahedron) from the
    DFT$+$U calculation. Solid lines and dashed lines represent
    up-spin and down-spin bands, respectively.}
  \label{fig2}
\end{figure}

\begin{figure}
   \includegraphics[width=8.5cm]{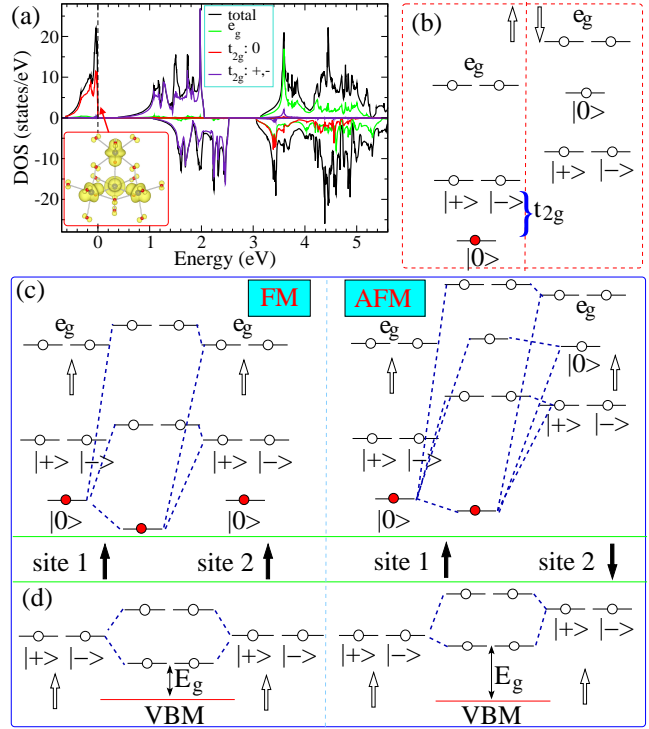}
  \caption{(color online) (a) The total and partial density of
    states of  the FM state from the DFT$+$U calculation. The inset
    shows the density distribution associated with the occupied V 3$d^1$
    states. (b) A schematic illustration of electron configuration of
    the V$^{4+}$ ion.
    (c) The interactions between the 3d-states of two adjacent
    V$^{4+}$ ions when
    their spins have the FM and AFM arrangements. Only the hopping
    processes which lead to the overall energy lowering of the up-spin
    $a_{1g}$ state of the spin site $1$ are shown.
    (d) The interactions between the 3d-states of two adjacent
    V$^{4+}$ ions leading to the
    CBM positions when their spins have the FM and AFM arrangements.  }
  \label{fig3}
\end{figure}

The band structure and density of states calculated for the
ferromagnetic (FM) state of Y$_2$V$_2$O$_7$ by the DFT$+$U method are
shown in Fig.~\ref{fig2}(a) and Fig.~\ref{fig3}(a), respectively. Both the valence
band maximum (VBM) and conduction band minimum (CBM) are majority-spin states.
There is an indirect band gap of about 0.33 eV between the VBM state
near the $W$ point and the CBM state at the
$\Gamma$ point. This band gap is consistent with the
experimentally measured activation energy 0.2 eV \cite{Shamoto2001}.
Our calculations show that the FM state of Y$_2$V$_2$O$_7$ has a nonzero band gap
when $U-J \ge  1.5$ eV. 
The four states between $-0.7$ eV and 0 eV (with zero set at the VBM level) are the occupied 
3$d$ states of the four V$^{4+}$ ions per unit cell. The electron density associated with
the four states, plotted in the inset of Fig.~\ref{fig3}(a),
clearly shows that these occupied orbitals are the $a_{1g}$ states, in
agreement with experiment \cite{Kiyama2006,Ichikawa2005}.
The analysis of the partial density of states provides further insight
into the electronic structures of Y$_2$V$_2$O$_7$.
For the spin majority part, the orbital levels are consistent with
the trigonal crystal field splitting.
For the spin minority part, 
the empty $|+\rangle$ and $|-\rangle$ states have lower energies than the
unoccupied $|0\rangle$ state.
This is so because the intra-orbital onsite Coulomb interaction $U$ is much
larger than
% 
% The wording  “the sum of” should be deleted!
%
the interorbital onsite Coulomb interaction ($U'$),
and the trigonal crystal field splitting  between the $a_{1g}$ and
$e'_{g}$ states. The spin down $a_{1g}$  state
lies slightly lower in energy than the spin down $e_g$ states, and both states are
delocalized with strong hybridization between them. The electron configuration of the
V$^{4+}$ ion is shown schematically in Fig.~\ref{fig3}(b).

The symmetric spin exchange parameters were extracted by mapping the
energy differences between ordered collinear spin states
obtained from DFT$+$U calculations onto the corresponding
energy differences obtained from the quantum Heisenberg Hamiltonian
for the spin-$1/2$ system: $H=
\sum_{i<j} J_{ij} \hat{\mathbf S_i}\cdot \hat{\mathbf S_j}$. We consider all
symmetric spin exchange interactions up to the third nearest
neighboring pairs [see Fig.~\ref{fig1}(b)];
the NN exchange $J_1$ within each V$_4$ tetrahedron and the next NN exchanges $J_2$,
$J_3$ and $J_4$. To evaluate these four spin exchange parameters reliably,
we considered 33 different ordered spin states and then determined them
by performing a linear
least-square fitting analysis \cite{Xiang2009}. Our calculations show that $J_1= -7.09$
meV, $J_2= -0.07$ meV, $J_3 = -0.31 $ meV, and $J_4 = -0.28 $ meV,
 namely, $J_2$, $J_3$ and $J_4$ are negligibly small compared with the NN
 FM exchange $J_1$. Using the calculated $J_1$, we estimate the Curie-Weiss  temperature
$\theta =z S(S+1) J_1 /3k_B   = 6 \times \frac{1}{2}(\frac{1}{2}+1) J_1 /3k_B = 122$ K,
which is close to the observed value 106 K for Lu$_2$V$_2$O$_7$ \cite{Zhou2008}.
The calculated spin exchange parameters show that the magnetic ground state is the FM
state, in agreement with experiment.
Interestingly, we find that $J_1$ is always ferromagnetic for any
reasonable $U$ and $J$ values ($U<8$ and $J \ge 0$), which is not in support of
the previous prediction \cite{Ichikawa2005} that antiferromagnetism is
favorable when $J<0.7$.

We now discuss the microscopic origin of the ferromagnetism in R$_2$V$_2$O$_7$ by
comparing the electron hopping processes between adjacent spin sites
$1$ and $2$ for cases when they have the FM and AFM arrangements. By symmetry, 
the overall total energy gain by the hopping is twice that of the majority-spin
$a_{1g}$ state of site 1 ($a_{1g}^{1\uparrow}$).
 In the FM case, the  $a_{1g}^{1\uparrow}$ state can hop to the
majority-spin  $e'_g$ and $e_g$ states but not $a_{1g}$ of site 2 according to the 
Pauli exclusion principle. In the AFM case, the $a_{1g}^{1\uparrow}$
can hop to all minority-spin states of site 2.
The energy gain from the virtual hopping is larger for the
FM case than for the AFM case because the empty up-spin states of site 2
are closer in energy to the filled $a_{1g}^{1\uparrow}$ state for the
FM case and the hopping between $a_{1g}^{1\uparrow}$ state and
the minority-spin $a_{1g}$ state of site 2 is negligible due to our finding that the
minority-spin $a_{1g}$ state lies much higher in energy
than the minority-spin $|+\rangle$ and $|-\rangle$ and a rather
small transfer integral between $a_{1g}$ states.

Fig.~\ref{fig2}(b) shows the band structure of an AFM state in which
there are two up- and two down-spins in every V$_4$ tetrahedron [see inset of
Fig.~\ref{fig2}(b)]. As in the FM case, there is an indirect
band gap between the VBM state
near the $W$ point and the CBM state at the
$\Gamma$ point. However, there is an important difference: The band
gap of the AFM state is about 0.62 eV, which is almost twice that
of the FM state. By taking the level of the Y $4s$ semicore level as
the reference, we find that the VBM level of the FM state is at almost
the same (only about 0.01 eV higher) as that of the AFM state.
Therefore, the CBM of the AFM state is much higher than that of the FM
state. The reason why the AFM state has a higher CBM and thus a larger
band gap is illustrated in Fig.~\ref{fig3}(d). When
the spins of two neighboring V ions have an FM coupling, the up-spin
$e'_{g}$ states of the adjacent sites have
the same energies so that the lowest energy state
is lower than the $e'_{g}$ level by $t'$, where $t'$ is the hopping
integral between the adjacent $e'_{g}$ states. In the AFM case, however, 
the lowest energy state is lower than the majority-spin $e'_{g}$ level by
$t'^2/\Delta$, where $\Delta$ is the exchange splitting and $\Delta > t'$.
Therefore, the AFM state has a higher CBM and thus a larger band gap.
This fact naturally explains the negative MR
observed just above the ferromagnetic Curie temperature $T_c$. When the
temperature is lowered towards $T_C$, the
spins have a tendency to order ferromagnetically, but are not fully
aligned. The application of an external magnetic field (about 5 T)
will help align the spins ferromagnetically. Thus, the band gap decreases with increasing the
magnetization, so that the resistivity of R$_2$V$_2$O$_7$ would be reduced by
external magnetic field hence leading to the negative MR
effect. This explanation is consistent with the observation that the
maximum MR effect occurs at 75 K, right above the Curie
temperature [See Fig. 3(a) of Ref. \cite{Zhou2008}].
We note that this new mechanism of negative MR should
be also applicable to other ferromagnetic insulators (e.g.,  EuO).

%D parameter and anisotropy
Let us now examine the magnetic properties that require the
consideration of SOC. For two interacting spins, the 
SOC can induce DM
antisymmetric interactions $H_{DM}=  {\mathbf D_{ij}} \cdot (\hat
{\mathbf S_i} \times \hat {\mathbf S_j})$ (${\mathbf D_{ij}} = D
{\mathbf d_{ij}}$, $D=|{\mathbf D_{ij}}|$ and ${\mathbf d_{ij}}$ is a unit vector).
According to the crystal symmetry, the DM vector
for a V-V edge of each V$_4$ tetrahedron is perpendicular to the V-V
bond and is parallel to the opposite edge of the V$_4$ tetrahedron [Fig.~\ref{fig1}(d)].
To evaluate the magnitude of the DM interaction term D, we consider
two spin configurations shown in Fig.~\ref{fig1}(e).
In one spin configuration, the four spins are aligned along the
$X$, $Y$, $Z$, and $Z$ axes, respectively [see Fig.~\ref{fig1}(a) for  the
definition of the global coordinate system $XYZ$]. From this configuration, we
generate the other spin configuration by performing a reflection
operation of each spin with respect to the $XZ$ plane containing the spin site. The only
difference from the first configuration is that the spin at the second site now points along the
$-Y$ direction. It can be easily shown that the two spin configurations
have the same spin exchange interactions and the same single-ion anisotropy interactions (see
below). In terms of the quantum expression of the DM interaction Hamiltonian,
it is found that the total DM interaction vanishes for the first spin configuration, but is
$D\sqrt{2}/2$ per V$_4$ tetrahedron for the second spin configuration. By using
the energy difference between the two spin
configurations obtained from the DFT$+$U$+$SOC calculations, $D$ is
estimated to be $0.34$ meV. The $D$ value is rather insensitive to the
calculation parameters ($U$ and $J$). Thus, $D$ is of the order of 5\%
of the NN spin exchange $J_1 = -7.09$ meV, i.e., $|D/J_1|=0.048$, which is
almost an order of magnitude smaller than
the value $D/J_1 = 0.32$ deduced by Onose {\it et al.} from analyzing the
magnetic field dependence of the thermal Hall conductivity in terms of
their model for the magnon Hall effect. This raises a serious question as 
to whether the observed thermal Hall
effect can be described solely in terms of the DM interaction.

%\clearpage

Another important consequence of the SOC interaction is that the magnetic
moment of each spin site gets a preferred orientation in space
with respect to the crystal lattice. However, it is commonly believed that a spin-$1/2$
ion has no single-ion anisotropy because the spin doublet state is not
split by the zero-field splitting term $CS_z^2=C I/4 $ for $S=1/2$ ($C$ is a constant and $I$
is a $2\times 2$ unit matrix) \cite{Dai2008}.
Here, we find that the spin-$1/2$ V ions of R$_2$V$_2$O$_7$ have 
substantial single-ion anisotropy, which is not described by the usual zero-field splitting term.
The single-ion anisotropy of the V$^{4+}$ ion can be estimated in
two ways. First, we replace three of the four V$^{4+}$ ions in each V$_4$
tetrahedron of Y$_2$V$_2$O$_7$ with non-magnetic Ti$^{4+}$ Ti$^{4+}$ ions to obtain
Y$_2$Ti$_{\frac{3}{2}}$V$_{\frac{1}{2}}$O$_7$, which has no NN pairs
of V$^{4+}$ ions. Our DFT$+$U$+$SOC calculations show that
the easy axis of the spin-$1/2$ V ion is along the three-fold rotational
axis [$z'$ in Fig.~\ref{fig1}(c)] of the distorted VO$_6$ octahedron.
The state with spin moments parallel to the $z'$ axis
is more stable than that with the spin moments perpendicular to the
$z'$-axis by about $0.81$ meV per V.
Second, we consider two spin configurations to obtain a more accurate
value of the single-ion anisotropy. In one configuration, all spins are along
the easy axis directions such that, in a V$_4$ tetrahedron, one spin is pointed out from the
center and the remaining three spins are pointed to the center. From this spin
configuration, we obtain the other spin configuration by rotating the directions of
all the spins around the global $Z$ axis by 120$^\circ$. In the resulting spin
configuration all spins are perpendicular to the easy axis directions.
The two spin configurations are the same in the symmetric exchange
interactions and in the DM interactions.
Our DFT$+$U$+$SOC calculations for the two spin configurations show
that the spin orientation along the easy axis is more stable than that along the
hard axis by $0.91$ meV per V.
Therefore, it is unequivocal that a spin-$1/2$ ion can have
significant single-ion anisotropy, contrary to the general belief. 
It should be noted that single-ion
anisotropy for a spin-$1/2$ ion is not excluded according to the general expression of
the $\mathbf{L}\cdot \mathbf{S}$ Hamiltonian \cite{Dai2008,Xiang2008}.
In the traditional effective spin Hamiltonian approach \cite{Dai2008,Majlis2000},
the energy for an orbital/spin basis function $|LM_L\rangle|SM_S\rangle$ does not
depend on the spin projection $| M_S\rangle$, whereas
an exchange splitting is always present in any realistic magnetic system,
as found in our first principles calculations. 
%
% Please remove the following
%The single-ion anisotropy for $S=1/2$
%we found in this work might be also important for the Cu$^{2+}$ ($S=1/2$)
%ions in cuprate superconductors.
%

As described above, the single-ion anisotropy energy is
much greater than the DM interaction parameter in R$_2$V$_2$O$_7$. We
now estimate how the DM interaction parameter is affected when the single-ion anisotropy
energy is neglected. The single-ion anisotropy for the spin-$1/2$ V$^{4+}$ ion cannot be
described by any quantum spin Hamiltonian. Our calculations show that the
single-ion anisotropy energy can nevertheless be written as
$H_{ani}= A \sum_i (\mathbf{S}_i \cdot
\mathbf{z}'_{i})^2 $ ($|\mathbf{S}|=1/2$ and $A=-3.64$ meV) when the
spins are treated as classical vectors. Then, the total classical spin Hamiltonian for
R$_2$V$_2$O$_7$ is expressed as:
\begin{equation}
H=\sum_{i<j} J {\mathbf S}_i \cdot {\mathbf S}_j + D \sum_{i<j} {\mathbf
  d}_{ij} 
\cdot ({\mathbf S}_i \times {\mathbf S}_j)+A \sum_i (\mathbf{S}_i
\cdot
\mathbf{z}'_{i})^2.
\end{equation}
Using the spin Hamiltonian, our Monte Carlo simulation shows that the spin ground 
state is nearly FM
with the moments aligned along the [001] direction, which is in
agreement with the slight anisotropy displayed in the magnetization curve \cite{Onose2010}.
If we renormalize the single ion anisotropy term into the  DM
interaction term, the Hamiltonian is rewritten as:
$H=\sum_{i<j} J {\mathbf S}_i \cdot {\mathbf S}_j + D^{eff} \sum_{i<j}
{\mathbf  d}_{ij} \cdot ({\mathbf S}_i \times {\mathbf S}_j)$, where $D^{eff}$ is the
effective DM interaction parameter to be obtained by neglecting the
single-ion anisotropy. For two spin configurations that are related to each other by a
mirror-plane symmetry [e.g., see Fig.~\ref{fig1}(e)], the energy differene
can be used to extract   $D^{eff}$. We generate several random spin configurations
to find that $\frac{D^{eff}}{D}$ can be as large as 20. This shows
that the effective DM interaction can be much larger than the actual
DM interaction when the single-ion anisotropy is neglected.

In summary, R$_2$V$_2$O$_7$ exhibits negative MR because its 
band gap depends on the spin arrangement, with the smallest gap for the FM state. 
The $V^{4+}$ ions of R$_2$V$_2$O$_7$ exhibits an easy-axis single-ion 
anisotropy that is much stronger than the DM interaction term $D$, despite the common belief that 
spin-$1/2$ have no single-ion anisotropy. The effective $D$ value 
evaluated can be unreasonably large when this anisotropy is neglected, as found 
from the analysis of the magnon quantum Hall effect of Lu$_2$V$_2$O$_7$. Thus,
the consideration of the single-ion anisotropy is necessary to
formulate a more complete theory for the observed magnon Hall effect.

%%Acknowledgements
Work at Fudan was partially supported by NSFC,
Pujiang plan, and Program for Professor of Special Appointment
(Eastern Scholar). Work at NREL was supported by U.S. DOE under
Contract No. DE-AC36-08GO28308, and that at NCSU U.S. DOE, under
 Grant No. DE-FG02-86ER45259.
%\clearpage

\clearpage

%\clearpage

\end{document}